\documentclass[10pt,letterpaper]{article}
\usepackage[top=0.85in,left=2.75in,footskip=0.75in,marginparwidth=2in]{geometry}

\usepackage[utf8]{inputenc}

\usepackage{cite}

\usepackage{nameref,hyperref}

\usepackage[right]{lineno}

\usepackage{microtype}
\DisableLigatures[f]{encoding = *, family = * }

\raggedright
\usepackage{changepage}

\usepackage[aboveskip=1pt,labelfont=bf,labelsep=period,singlelinecheck=off]{caption}

\makeatletter
\renewcommand{\@biblabel}[1]{\quad#1.}
\makeatother

\usepackage{lastpage,fancyhdr,graphicx}
\usepackage{epstopdf}
\fancyheadoffset[L]{2.25in}
\fancyfootoffset[L]{2.25in}

\usepackage{color}

\definecolor{Gray}{gray}{.25}

\usepackage{graphicx}

\usepackage{sidecap}

\usepackage{wrapfig}
\usepackage[pscoord]{eso-pic}
\usepackage[fulladjust]{marginnote}
\reversemarginpar

\begin{document}
\vspace*{0.35in}

\begin{flushleft}
{\Large
\textbf\newline{Vortex driven Schwinger pair creation in the magnetosphere of SgrA*}
}
\newline
\\
Zaza N. Osmanov 
\\
\bigskip
School of Physics, Free University of Tbilisi, 0183, Tbilisi, Georgia
\\
E. Kharadze Georgian National Astrophysical Observatory, Abastumani 0301, Georgia
\\
\bigskip
z.osmanov@freeuni.edu.ge

\end{flushleft}

\section*{Abstract}
In this work, we explore the possibility of Schwinger pair creation triggered by magneto-centrifugal effects in the magnetosphere of Sgr A*. We show that these effects become extremely efficient in the presence of a vortex-driven magnetic field, whose strength exceeds previous estimates by several orders of magnitude. The dynamics of magneto-centrifugally accelerated charged particles leads to charge separation, thereby parametrically exciting Langmuir waves. The associated electric field grows exponentially and, upon reaching the Schwinger critical threshold, initiates efficient electron–positron pair production, which is ultimately saturated by annihilation processes.


\section{Introduction}

Recently, interest in Sgr A* has increased substantially, largely motivated by the first images of the event-horizon–scale structure of the Galactic center black hole obtained by the Event Horizon Telescope \cite{EHT}. Studies of this source span a broad range of topics, including the dynamics of its surrounding environment \cite{stars}, the origin of flaring activity \cite{flare}, and potential connections to dark matter \cite{DM}.

It is evident that many of the processes occurring in the immediate vicinity of Sgr A* are largely governed by the properties of its magnetosphere, making its detailed study particularly important. For instance, a wide range of phenomena in the magnetospheres of supermassive black holes are thought to be controlled by the presence of electron–positron pairs. It is generally accepted that two primary mechanisms are responsible for their production. In the first scenario, high-energy photons interact with softer photons originating in the accretion disk, leading to the creation of electron–positron pairs \cite{pair1}. According to the Penrose mechanism, photons with energies of order MeV produced in the inner regions of the accretion disk can gain additional energy upon entering the ergosphere. If amplified to GeV energies, such photons may subsequently produce electron–positron pairs through interactions with protons \cite{pair2}.

Recently, it has been shown that the mechanisms discussed above are not the only channels through which electron–positron pairs can be produced. In particular, Refs.\cite{PCpulsar,PCAGN} investigate a centrifugally driven Schwinger mechanism operating in the magnetospheres of pulsars and active galactic nuclei, demonstrating that the resulting particle number densities can substantially exceed those predicted by conventional mechanisms. The essence of this process is as follows: a strong magnetic field in the magnetosphere of a Kerr-type black hole or a pulsar constrains charged particles to move along rotating magnetic field lines. This combination leads to magneto-centrifugal acceleration, which can boost particles to very high energies (VHE). Ultimately, magneto-centrifugal effects give rise to an exponentially growing charge-separation process, during which the electrostatic field may reach the so-called Schwinger threshold $E_S = \pi m_e^2c^3/(e\hbar)\simeq 1.4\times 10^{14}$ statvolt cm$^{-1}$ \cite{heisen,sauter,schwinger} ($m_e$ and $e$ represent the electron's mass and charge respectively, $c$ is the speed of light and $\hbar$ denotes the Planck's constant).

The magneto-centrifugal acceleration mechanism plays a crucial role in driving the Schwinger process. A significance of the mentioned acceleration process first emphasized in the seminal work of Gold \cite{gold} and later applied to black holes by Blandford and Znajek \cite{BZ}. In applications to black hole magnetospheres \cite{bodo,blazar}, this acceleration mechanism has been shown to account for particle energies reaching the TeV range. In addition to particle acceleration, magneto-centrifugal effects have also been demonstrated to efficiently excite Langmuir waves \cite{jet,zev,pev} with the efficiency of this process being strongly dependent on the strength of the magnetic field in the magnetosphere.

In \cite{doz}, a photon mass of order $10^{-15}$ eV was considered consistent with experimental upper limits derived within Higgs-type scenarios \cite{lakes,luo}. It was shown that the resulting massive-photon vortices can be captured by a black hole embedded in a photon sea, allowing magnetic fields to grow to extremely large strengths over cosmological timescales. In most astrophysical environments, however, black holes are surrounded by accreting matter, in which vortical structures arise naturally without the need to invoke a finite photon mass. When combined with a graviton-based description of the black hole, this process generically leads to the generation of extremely strong magnetic induction \cite{vortex}. Therefore, both scenarios can result in a nearly saturated magnetic field, $B$. In particular, the maximum magnetic field that a black hole can confine may be estimated using an equipartition argument, whereby the magnetic energy becomes comparable to the total energy of the black hole $B^2R_g^3\simeq Mc^2$ (dimensionless factors are omitted here), resulting in
 \begin{equation}
\label{mag} B\simeq\frac{c^4}{MG^{3/2}}\simeq 6\times 10^{12} G,
\end{equation}
where $R_g = GM/c^2$ is the gravitational radius, $M\simeq 4\times 10^6 M_{\odot}$ denotes the mass of SgrA* \cite{EHT} and $G$ is the gravitational constant. This value exceeds magnetic field strengths estimated via the equipartition approximation based on either the thermal luminosity of the accretion disk or the non-thermal emission. Although the Schwinger mechanism has previously been considered in the context of active galactic nuclei \cite{PCAGN}, the aim of the present study is different: (a) we focus on the magnetosphere of Sgr A*, and (b) we explore a vortex-driven magnetic field that is several orders of magnitude stronger than the fields generated by the standard mechanisms discussed in earlier works.

Therefore, the study of pair creation also under the condition of vortex driven magnetic fields could be significant.

The paper is organized as follows: in section 2, we discuss the mechanism of the centrifugally driven pair creation, apply it to SgrA* and obtain results and in sections 3, we summarise them.

\begin{figure}
  \centering {\includegraphics[width=11cm]{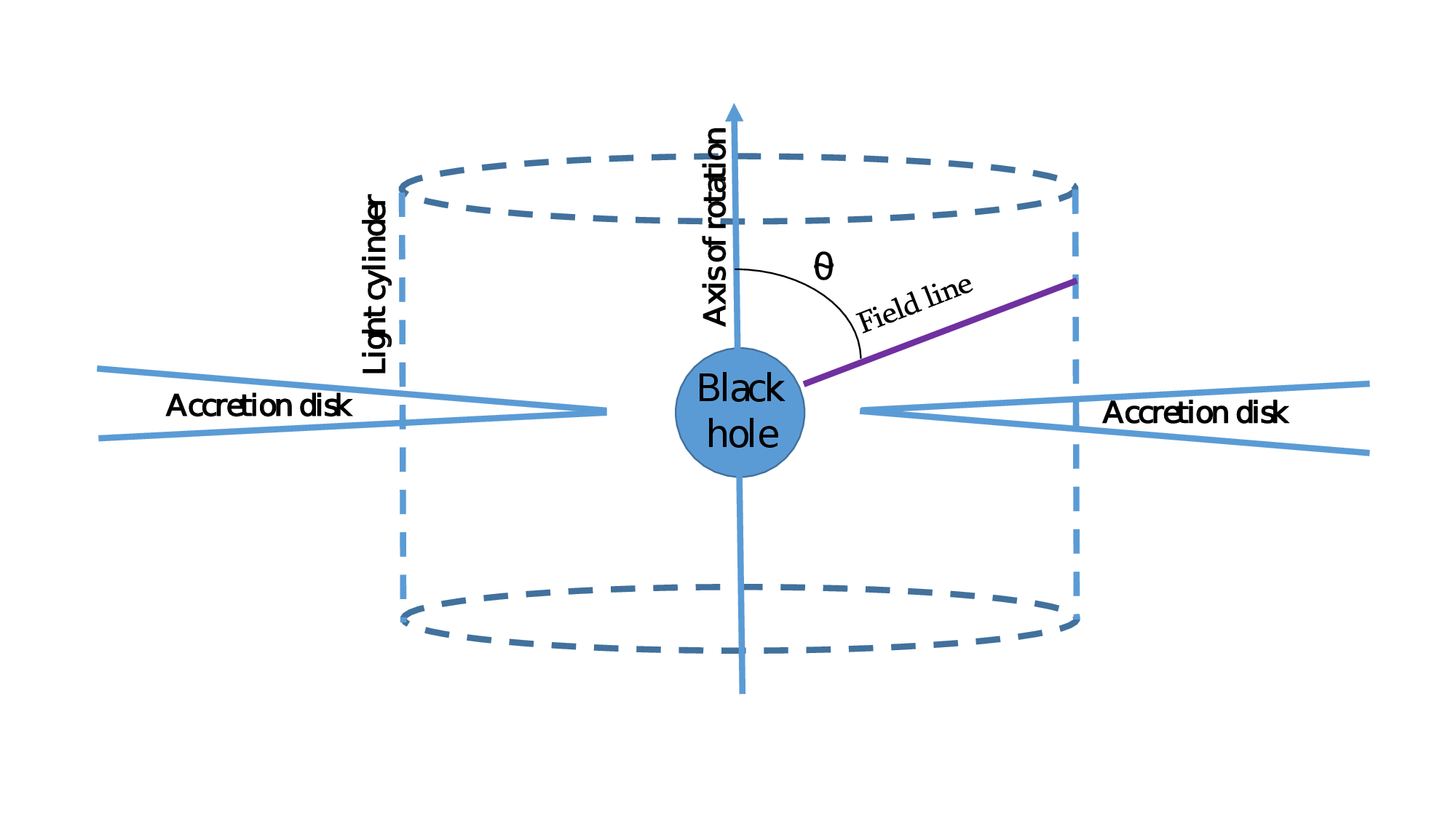}}
  \caption{Sketch of the model, with the centrifugally accelerated co-rotating particles moving along straight magnetic field lines
in the nearby zone of the LC area .}\label{fig1}
\end{figure}

\section{Discussion and results}
It is straightforward to show that, even for the conventional magnetic field \cite{bodo},
 \begin{equation}
\label{mag0} 
B\simeq\left(\frac{2L}{r^2c}\right)^{1/2}\simeq 10\times\left(\frac{R_g}{r}\right)^{1/3}\; G,
\end{equation}
 which is much weaker than the vortex-driven magnetic field (see Eq. (\ref{mag})), relativistic protons remain closely tied to the magnetic field lines. For instance, a proton with a Lorentz factor $\gamma = 10^{4}$ has a Larmor radius $\gamma m_pc/eB\simeq 0.1$ cm many orders of magnitude smaller than the spatial scale of the black hole, $R_g \simeq 6\times 10^{11}$ cm. Here $L\simeq 5\times 10^{35}$ erg/s is the average luminosity of Sgr A* \cite{sgrlum}. For electrons, the gyro-radius is even smaller, and in the presence of a vortex-driven magnetic field, the corresponding scale decreases by several additional orders of magnitude. Therefore, the frozen-in condition for the plasma is a valid and realistic approximation in the magnetosphere under consideration. 

This regime, when combined with rotation, results in efficient magneto-centrifugal acceleration and the subsequent pair-creation process. Assuming that SgrA* is a Kerr-type black hole, the corresponding angular velocity can be expressed as \cite{kerr,bardeen}
\begin{equation}
\label{rotat} \Omega\approx\frac{a c^3}{2GM\left(1+\sqrt{1-a^2}\right)}\approx
9.3\times 10^{-3} rad \;sec^{-2},
\end{equation}
where $a\simeq 0.65$ \cite{dokuch} is a dimensionless spin parameter of the black hole. It should be noted that the value of this parameter is the subject of considerable debate. In a recent study, three different values were considered, $0.95; 0.5; 0$ \cite{EHT}; Therefore, we adopt an intermediate value that is consistent with observational measurements and lies within the range discussed above. Moreover, the angular velocity is only weakly sensitive to the precise choice of this parameter.

In the vicinity of the light-cylinder (LC) region \cite{jet,zev,pev}—a hypothetical zone where the linear rotational velocity equals the speed of light (see Fig. 1)—this process is, however, limited by several factors. In the framework of this study, we assume rectilinear magnetic field lines. When moving in a strong magnetic field, relativistic particles lose energy efficiently through synchrotron radiation and rapidly transition to the ground Landau state, continuing to follow the co-rotating field lines. The particles then experience an effective reaction force; if this force exceeds the Lorentz magnetic force, the particles detach from the field lines, terminating acceleration and setting the maximum attainable Lorentz factor. This phenomenon is known as the breakdown of the bead-on-the-wire (BBW) approximation \cite{bodo}
\begin{equation}
\label{BBW1} 
\gamma_{max}^{BBW}\simeq A_1+\left[A_2+\left(A_2^2-A_1^6\right)^{1/2}\right]^{1/3}+\left[A_2-\left(A_2^2-A_1^6\right)^{1/2}\right]^{1/3},
\end{equation}
with
\begin{equation}
\label{A1} A_1 = -\frac{\gamma_0ctg^2\theta}{12},
\end{equation}
\begin{equation}
\label{A2} A_2 = \frac{\gamma_0e^2B^2}{4m^2c^2\Omega^2}+A_1^3,
\end{equation}
where $\theta$ denotes an inclination angle of the magnetic field (see Fig. 1).

A highly efficient limiting mechanism is the inverse Compton (IC) process, which occurs in a medium with a high density of photons. When accelerated particles encounter relatively soft photons—originating from accretion disks or other sources—they scatter off them, losing energy. As in the case described above, once an energy balance is reached, the particles attain a corresponding maximum Lorentz factor \cite{bodo}
\begin{equation}
\label{IC} 
\gamma_{max}^{IC}\simeq\left(\frac{8\pi mc^4}{\gamma_0\sigma_TL\Omega}\right)^2,
\end{equation}
where $\sigma_T$ is the Thomson cross section. It is worth noting that IC scattering of protons is strongly suppressed \cite{ahar}.

In general, charged particles move along curved trajectories, producing the so-called curvature radiation, characterized by the power, $P_c = 2e^2c\gamma^4/(3\rho^2)$ \cite{review}, which ultimately determines the maximum attainable Lorentz factor
\begin{equation}
\label{gcur} 
\gamma_{max}^{cur}\simeq\frac{1}{\gamma_0^{1/5}}\times\left(\frac{3mc^3\sin\theta}{e^2\Omega}\right)^{2/5}\times\left(\frac{\rho}{R_{lc}}\right)^{4/5},
\end{equation} 
where $\rho$ denotes the curvature radius of the trajectory and $R_{lc}=c/\Omega$ is the radius of the LC surface.

It is clear that the mechanism which provides the smallest value of the Lorentz factor, corresponds to the dominant mechanism limiting the acceleration process. For the conventional magnetic field (see Eq. (\ref{mag0})), electrons reach maximum Lorentz factors of order $1.8\times 10^6$ while protons are accelerated up to about $1.2\times 10^4$ with the process terminated by the breakdown of the bead-on-the-wire (BBW) approximation in both cases. In the case of a vortex-driven magnetic field (Eq. (\ref{mag})), the maximum attainable Lorentz factor for electrons is of order $1.6\times 10^{10}$, with curvature radiation being the primary limiting mechanism, whereas protons can reach Lorentz factors of approximately, $2.8\times 10^{11}$, limited by the BBW. These considerably higher values are natural, as stronger magnetic fields enhance the efficiency of centrifugal acceleration.
In the magnetosphere, one expects a distribution of particle energies. In this work, we adopt the equipartition approximation, assuming that the energy density is equal across different species $\gamma_{max}n_{_{GJ}}\simeq\gamma n$, where $n_{_{GJ}} = {\bf \Omega \cdot B}/(2\pi ec)$, is the Goldreich-Julian density \cite{GJ}, which is a density in rotational induced media.

This means that the charge separation phenomenon depending on the efficiency of the centrifugal mechanism and the resulting parametrically induced exponentially growing electrostatic field is quite realistic. In particular, in \cite{jet} it has been shown that the growth rate, $\Gamma$, of the corresponding parametric process (electric field grows as $e^{\Gamma t}$), being very efficient on the LC zone is given by
\begin{equation}
 \label{grow1}
 \Gamma= \frac{\sqrt3}{2}\left (\frac{\omega_1 {\omega_2}^2}{2}\right)^{\frac{1}{3}}
 {J_{\mu}(b)}^{\frac{2}{3}},
\end{equation}
where $\omega_{1,2}\equiv\sqrt{4\pi e^2n_{1,2}/m_{1,2}\gamma_{1,2}^3}$ is the plasma frequency of a corresponding specie (electrons and protons), $n_{1,2}$, $m_{1,2}$ and $\gamma_{1,2}$ are respectively the density, mass, and the Lorentz factor of the corresponding components, $J_{\nu}(x)$ denotes the Bessel function of the first kind, $b\simeq2ck/\Omega$ and $\mu = \omega_e/\Omega$. 

Considering a conventional magnetic field and two particle species with Lorentz factors $\gamma_1 =10$ (protons) and $\gamma_2 = 100$ (electrons), it can be shown that in the equatorial plane ($\theta = 90^\circ$) the instability time-scale, $\tau = 1/\Gamma \simeq 0.2$ s, is many orders of magnitude shorter than the kinematic (escape) time-scale of the rotating magnetosphere, $P = 2\pi/\Omega \simeq 670$ s. 

Similarly, for a vortex-driven magnetic field and field-line inclination angles $\theta = 1^\circ - 3^\circ$, with $\gamma_1 = 10$ (electrons) and $\gamma_2 = 100$ (protons), the instability time-scale $\tau \simeq 5 \times 10^{-4}$ s is significantly shorter than in the conventional field case. Our goal was to consider relativistic factors that can be reached without difficulty, in particular, achieving ultra-high energies is typically the main challenge, and it is generally believed that particles with the Lorentz factors mentioned above dominate in magnetospheric astrophysical plasmas of black holes. For larger inclination angles, Landau damping becomes dominant and suppresses the instability. It can also be readily shown that, within this same angular range, the instability time-scale is much shorter for the vortex-driven magnetic field than for the conventional magnetic field.

Thus, during the amplification process, the exponentially growing electric field, $E\simeq E_0\exp(\Gamma t)$, will eventually reach the Schwinger threshold, leading to efficient pair production at a rate given by \cite{schwinger,MP}
\begin{equation}
 \label{rate}
 R\equiv\frac{dN}{dtdV} = \frac{e^2E^2}{4\pi^3c\hbar^2}\sum_{k}\frac{1}{k^2}exp\left({-\frac{\pi m^2c^3}{e\hbar E}k}\right).
\end{equation}
For the initial value of the electrostatic field, $E_0$, one can use Gauss's law, $E_0\simeq 4\pi n\Delta r$ and $\Delta r\simeq\gamma_0R_{lc}/(2\gamma)$ \cite{review}.

This process occurs very rapidly, causing the energy density of the electrostatic field to develop exponentially. Simultaneously, as the field approaches the Schwinger threshold, efficient pair production begins, and the process naturally reaches saturation when the power density of the pair plasma equals that of the electric field \cite{PCAGN}
\begin{equation}
\label{term1} 
2m_ec^2R(t)\simeq\frac{d}{dt}\left(\frac{E^2(t)}{8\pi}\right).
\end{equation} 
One can numerically estimate the time-scales, $t_0$,at which this saturation condition is met in both cases. For a conventional magnetic field, the resulting number density of produced pairs, $R(t_0)t_0$, reaches $\sim 10^{30}$ cm$^{-3}$, a similar situation occurs in the vortex-driven magnetic field. Such an extremely high number density significantly increases the probability of annihilation, which appears to become the dominant saturation mechanism, determining the annihilation rate \cite{ann,PCAGN} 
\begin{equation}
\label{ann} 
\Lambda\simeq 2\pi c r_e^2 n^2,
\end{equation} 
and the production rate, which are of the same orders of magnitude \cite{PCAGN}
\begin{equation}
\label{bal} 
R(\tau)\simeq\Lambda\simeq 2\pi c r_e^2 \left(\int_0^{\tau}R(t)dt\right)^2.
\end{equation} 
Taking the derivative of Eq. (22) with respect to $\tau$ and neglecting the term $e^{\Gamma\tau}$  relative to $e^{2\Gamma\tau}$ on the left-hand side of the equation straightforwardly yields an estimate for the electron and positron number densities
\begin{equation}
\label{dens} 
n\simeq\frac{\Gamma}{2 \pi c r_e^2},
\end{equation} 
which for the conventional magnetic field and the vortex-driven magnetic field for $\theta = 1^0$ equals $3\times 10^{14}cm^{-3}$ and $3\times 10^{18}cm^{-3}$ respectively. As is clear, the super-strong magnetic field guaranties very high densities of particles. By combining Eqs. (\ref{bal},\ref{dens}) one can estimate the pair creation rate, which for the same parameters equals $1.2\times 10^{23}cm^{-3} s^{-1}$.

The results suggest that near the rotation axis of Sgr A*, an extremely high concentration of electron-positron pairs is likely to accumulate, naturally enhancing the annihilation process. Consequently, one can expect Doppler-shifted annihilation lines originating from this region. Specifically, the emitted radiation should experience both red and blue shifts, resulting in a detectable energy range approximately spanning $(MeV/\gamma - \gamma\; MeV)$. For instance, for a Lorentz factor on the order of $10$, the corresponding energy interval is approximately $100 keV - 10 MeV$. 

However, the total luminosity of the process strongly depends on the magnetic field topology and requires a dedicated analysis. It should also be noted that the resulting radiation is efficiently absorbed by cosmic dust \cite{carroll}, often complicating direct observations. Consequently, studying the outcomes of pair creation is a complex task and should be approached both through direct detection of the emission lines and via secondary effects, such as the efficient heating of the black hole magnetosphere.

\section{Conclusions}

It has been demonstrated that the central black hole of our galaxy, Sgr A*, acts as an efficient accelerator of particles via the magneto-centrifugal mechanism.

In addition to the conventional magnetic field generation mechanism, we have considered the vortex-driven magnetic field, which exceeds the conventional field by several orders of magnitude. It was shown that particles accelerated centrifugally in this environment induce an exponentially growing electric field.

Over time, the growing electric field may approach the Schwinger threshold, triggering efficient pair creation. The Schwinger mechanism can generate extremely high electron-positron number densities, producing annihilation emission lines that could potentially be detectable.

\section*{Acknowledgments}
The research was supported by Shota Rustaveli National Science Foundation of Georgia (SRNSFG) Grant: FR-23-18821 and a German DAAD scholarship within the program Research Stays for University Academics and Scientists, 2024 (ID: 57693448). ZO is grateful to prof. G. Dvali for fruitful discussions and comments. ZO acknowledges Max Planck Institute for Physics (Munich) for hospitality during the completion of this project.

\nolinenumbers





\end{document}